\documentstyle[epsf,12pt]{article}

\newcommand{\be}{\begin{equation}}
\newcommand{\ee}{\end{equation}}
\newcommand{\ba}{\begin{eqnarray}}
\newcommand{\ea}{\end{eqnarray}}
\newcommand{\baa}{\begin{eqnarray*}}
\newcommand{\eaa}{\end{eqnarray*}}
\newcommand{\bb}{\begin{thebibliography}}
\newcommand{\eb}{\end{thebibliography}}
\newcommand{\ci}[1]{\cite{#1}}

\newcommand{\lab}[1]{\label{#1}}
\newcommand{\re}[1]{(\ref{#1})}

\newcounter{my}
\newcommand{\he}%
   {\stepcounter{equation}\setcounter{my}%
   {\value{equation}}\setcounter{equation}0%
   }%
\newcommand{\she}%
   {\setcounter{equation}{\value{my}}%
    }%

\begin{document}%

\vspace*{5mm}

\begin{center}
{\bf SOLITON SOLUTIONS OF INTEGRABLE HIERARCHIES \\[2mm]
AND COULOMB PLASMAS}

\vspace{5mm}

{\large Igor Loutsenko$^*$ and Vyacheslav Spiridonov$^+$}

\vspace{4mm}
{$^*$ \it
     Department of Physics, Joseph Henry Laboratories, Jadwin Hall, \\
     Princeton University, P.O. Box 708,
     Princeton, NJ 08544-0708, USA; \\
 e-mail:  loutseni@feynman.princeton.edu} \\[1mm]

{$^+$ \it Bogoliubov Laboratory of Theoretical Physics, JINR,
Dubna, \\  Moscow region 141980, Russia;
e-mail: svp@thsun1.jinr.ru
}\\[5mm]
\end{center}

\begin{abstract}
Some direct relations between soliton solutions of integrable hierarchies and
thermodynamical quantities of the Coulomb plasmas on the plane
are revealed. We find that certain soliton solutions of the
Kadomtsev-Petviashvili (KP) and B-type KP (BKP) hierarchies describe
two-dimensional one or two component plasmas at special boundary conditions
and fixed temperatures. It is shown that different
reductions of integrable hierarchies describe one (two) component
plasmas or dipole gases on one-dimensional submanifolds embedded in
the two-dimensional space.
We demonstrate application of the methods of soliton theory to
statistical mechanics of such systems.
\end{abstract}


\section{Introduction}

Recently we have shown \ci{LS} that the grand partition
functions of some one-dimensional
lattice gas models or equivalent to them partition functions of some
Ising chains, coincide with the $N$-soliton
tau-functions of various hierarchies of integrable
nonlinear evolution
equations. The present paper is the third one in the series and it
comprises a detailed comparison of exactly solvable Coulomb plasma
models on lattices with integrable equations.
In extension of our previous considerations we discuss
statistical mechanics of the Coulomb (logarithmic interaction) 
gases on intrinsic two-dimensional
geometric figures and various one-dimensional
submanifolds of the plane. In this way we do not merely reinterpret
previously known results \ci{G}-\ci{FJT}, but
also reveal  a number of new exactly solvable models.
It is natural to expect that the connection with integrable
equations gives a clue to the classification of solvable (at fixed
temperatures) plasma models.

A classical Coulomb plasma is a system of charged particles interacting through
the Coulomb potential. In the Euclidean space $R^n$
the Coulomb potential is defined as a solution to the Poisson equation
\begin{equation}
\Delta V(r,r^\prime)= - \omega_n\delta(r-r^\prime),
\qquad \omega_n=\frac{2\pi^{n/2}}{\Gamma(n/2)}, \quad r\in R^n
\label{laplace}
\end{equation}
with $V(r,r^\prime)$ obeying certain boundary conditions.
For the two-dimensional plane without boundaries,
equation (\ref{laplace}) has the following solution
\begin{equation}
V(z,z^\prime)=-\ln |z-z^\prime|,
\label{log}
\end{equation}
where $ z=x+\i y$ and $\omega_2=2\pi$.
A system of particles forms a stable plasma if the
opposite valued charges do not recombine with each other forming a gas of
neutral molecules.

In general the location of plasma is constrained to some domain in
which case there is a nontrivial interaction with the boundaries of
this domain. In particular, the normal component of the electric field
\begin{equation}
{\cal E} = -\nabla V
\label{field}
\end{equation}
should vanish on the surface of an ideal dielectric
\begin{equation}
{\cal E}_n = 0,
\label{normal}
\end{equation}
while the tangent component of this field vanishes on the surface of an ideal
conductor
\begin{equation}
{\cal E}_t =0.
\label{tangent}
\end{equation}

A useful way of solving the Poisson equation with such boundary conditions
is provided by the method of images.
In the present work we consider systems where every charge has either
a finite number of
images created by the boundaries or boundary conditions create periodic
lattices of images.
In the case of a finite number of images the solution to (\ref{laplace})
is a finite sum of logarithmic potentials (\ref{log})
created by a charge and its images.

The energy of such systems of $N$ particles
has the following form
\begin{equation}
E_N=\sum_{1\le i<j\le N} Q_iQ_jV(z_i,z_j)+\sum_{1\le i\le N} Q_i^2v(z_i)
+\sum_{1\le i\le N} Q_i\Phi(z_i)
\label{energy}
\end{equation}
where $z_i=x_i+\i y_i$ and $Q_i$ denote the coordinate and the charge
of the $i$-th particle on the plane.

The first term in (\ref{energy}) is the energy of interaction between
different charges. The second term is the sum of self-energies.
This term is constant at certain
boundary conditions, e.g. when the plasma is homogeneous or if it is
restricted to a line with transverse boundary conditions.
In general it arises as the
energy of interaction between the charge and its own images.
The third term describes an interaction of charges with
external fields.

The grand partition function of a system of particles of $s$ different
species is
$$
G=\sum_{n_1=1}^{N_1}\dots\sum_{n_s=1}^{N_s}\frac{\zeta_1^{n_1}\dots\zeta_s^{n_s}}
{n_1!\dots n_s!}Z_{n_1\dots n_s}
$$
\begin{equation}
=\sum_{n_1=1}^{N_1}\dots\sum_{n_s=1}^{N_s}\frac{1}
{n_1!\dots n_s!}\int e^{-\Gamma H_{n_1\dots n_s}+\mu_1 n_1 + \dots +
\mu_s n_s} {\rm d} \Omega,
\label{grand}
\end{equation}
where $\mu_1,\dots,\mu_s $ denote chemical potentials
($\zeta_s=e^{\mu_s}$ are the fugacities)
and $\Omega$ is the integration measure
over the configuration space occupied by particles.
In (\ref{grand}) we have
introduced the dimensionless inverse temperature
\begin{equation}
\Gamma=\beta Q^2, \qquad \beta=1/kT
\label{gamma}
\end{equation}
and the dimensionless Hamiltonian
\begin{equation}
H_{n_1\dots n_s}=\frac{1}{2}\sum_{i\neq j} q_iq_jV(z_i,z_j)+\sum_{i} q_i^2v(z_i)
+\sum_{i} q_i\phi(z_i),
\label{hamiltonian}
\end{equation}
where $q_i=Q_i/Q$ and $\phi(z)=\Phi(z)/Q$.
For one and two component plasmas it is convenient to choose $Q=|Q_i|$,
so that the dimensionless charges $q_i=1$ in the one component
case and $q_i=\pm 1$ in the two component case.

The two-dimensional one and two component plasma models have been solved for
a variety of boundary conditions at the special value of the inverse temperature
$ \Gamma = 2 $ (see, e.g. \cite{G}-\cite{FJT} and references therein). In the
case of two component plasma an exact solution exists due to the well known
correspondence between this system and the free fermion point of the
Thirring/Sine-Gordon model. In technical terms the solution is possible
due to different determinant representations (the Cauchy determinant for the
two component plasma and the Vandermonde determinant for the one component
case). Such determinant representations allow one to solve models of
$\log$-gases on a line with the transverse
boundary conditions (e.g., see \cite{FJT}).
Note that the main part of the cited works deal with the continuous space
models, for an account of the lattice models, see, e.g.
\cite{LS,G,CJ}.
The literature on the Coulomb gases is enormous, many statistical
mechanics models have been related to them \ci{Nien}, there is a
relation to conformal field theories, etc. Still, the identification
of Coulomb plasmas on lattices with some boundaries and of
the famous multi-soliton systems
has been missed in the previous investigations.

\section{Basic observations}

Let us consider a lattice version of (\ref{grand}). We suppose that each
type of particles can occupy only a discrete set of points in the
complex plane. E.g., in the two component case the
positive and negative charge particles occupy sublattices $\{z_+\}$ and
$\{z_-\}$. We denote the union of all sublattices as $ \{z\} $.
No more than one particle is allowed at each site.
In this case, the integrals in (\ref{grand}) are replaced by discrete
sums over the lattice points and the whole partition function
can be rewritten in the following form \ci{B}
\begin{equation}
G=\sum_{\{\sigma\}}\exp\left(\frac{1}{2}
\sum_{z\neq z^\prime}W\left(z,z^\prime\right)\sigma(z)\sigma(z^\prime)
+\sum_{\{z\}}w(z)\sigma(z)\right),
\label{G}\end{equation}
where
\begin{equation}\label{W}
W\left(z,z^\prime\right)=-\Gamma q(z)q(z^\prime)V(z,z^\prime), \quad
w(z)=\mu(z)-\Gamma\left(q^2(z)v(z)+q(z)\phi(z)\right)
\end{equation}
and $\sigma(z)=0\;{\rm or}\; 1$ is an occupation number of
the site with the coordinate $z$.
The variables $q(z),\mu(z)$ are some functions of the lattice
coordinates. For example,
$q({z_\pm})={\pm 1}, \; \mu({z_\pm})=\mu_{\pm}$ for the
two component plasma.

Now, let us write out the $\tau$-function of  $N$-soliton solutions of
some integrable hierarchy in the Hirota form \cite{AS}
\begin{equation}
\tau_N=\sum_{\sigma=0,1}\exp\left(\frac{1}{2}\sum_{z\neq z^\prime}A_{z
z^\prime}\sigma(z)\sigma(z^\prime)+\sum_{\{z\}}
\theta(z)\sigma(z)\right),
\label{tau}\end{equation}
where the variable $z$ takes $N$ discrete values describing spectral
characteristics of solitons.
The function $\theta(z)$ parameterizes phases of solitons with
the index $z$ and $A_{zz^\prime}$ is the phase shift acquired as
a result of the scattering of solitons with the indices $z$ and
$z^\prime$ on each other.

Evidently, the expressions (\ref{G}) and (\ref{tau}) have the same form.
One just needs to make proper identifications between the phase shifts
$A_{zz^\prime}$ and the interaction potentials $W(z,z^\prime)$, and
between the phases $\theta(z)$ and the function $w(z)$.

Concluding this section we present brief explanations on the
structure of the expression (\ref{tau}).
One can track appearance of the $\tau$-function on the simplest
example of the Korteweg-de Vries (KdV) equation which emerges as a
compatibility condition of two linear problems:
\begin{equation}
L\psi(x)\equiv -\psi_{xx}(x)+u(x)\psi(x)=\lambda \psi(x),
\label{sch}
\end{equation}
\begin{equation}
\psi_t(x, t) = B \psi(x, t),
\qquad B \equiv -4\partial_x^3 + 6u(x,t)\partial_x+ 3u_x(x,t).
\label{time}\end{equation}
Eigenvalues of the Schr\"odinger operator (\ref{sch}) do not
depend on ``time" $t$: $ \partial \lambda /\partial t =0.$
As a result, the compatibility condition of \re{sch} and \re{time} takes the
following operator form: $ L_t=[B, L]$,
and yields the KdV equation
$$ u_t+u_{xxx}-6uu_x=0,  $$
describing the motion of shallow water.
Stable traveling wave solutions of this equation are called
solitons. The inverse scattering method allowed one to find
an explicit expression for the general $N$-soliton solution
\be
u(x, t)=- 2\partial_x^2 \ln \tau_N(x, t),
\lab{pottau}\ee
where $\tau_N$ is the determinant of some $N\times N$ matrix
called the tau-function. It can be written in the form
\begin{equation}
\tau_N = \sum_{\mu_i=0,1} \exp \left( \sum_{1\leq i<j \le N }
A_{ij} \mu_i \mu_j + \sum_{1\leq i\leq N} \theta_i\mu_i \right),
\label{iN_soliton}
\end{equation}
\begin{equation}\label{iphase}
\theta_i=k_ix-k_i^3 t +\theta_i^{(0)}, \qquad i, j=1, 2, \dots, N.
\end{equation}
The variable $k_i$ is related to the $i$-th eigenvalue of
the operator $L$ in the standard quantum mechanical meaning,
$\lambda_i=-k_i^2/4$, and it parameterizes the amplitude of
the $i$-th soliton which is proportional to $k_i^2$.
The scattering process is described by the time evolution
from $t\to -\infty$ to $t\to +\infty$. In this picture
$\theta_i^{(0)}/k_i$ are the zero time phases of solitons and
$k_i^2$ are equal to their velocities.

The phase shifts $A_{ij}$ describe relative retardation of $i$-th
soliton after the interaction with the $j$-th one. It is determined
by the formula
\begin{equation}
e^{A_{ij}}={ (k_i-k_j)^2 \over  (k_i+k_j)^2 }.
\label{iKDV_phase}
\end{equation}

In the standard approach $k_i$ are real. However,
formally one can take them as complex variables as well.
Then it is seen from (\ref{G}) that the plane coordinates
$z_i$ of some particular configuration of the Coulomb
particles (see below) play the role of $k_i$ and
correspond thus to complex amplitudes and velocities of solitons.
This leads to complexification of the
function $u(x,t)$ which requires somewhat different physical
interpretation of the KdV equation. However, for complex $k_i$
there is no complete coincidence of \re{G} and \re{tau} ---
the signs of modules in the Coulomb potential make $G$
\re{G} real and there are no such signs in \re{iKDV_phase}
and, so, $\tau_N$ is complex.
On the one hand, one may place the plasma onto the
$x$ or $y$ axis in order to reach the coincidence
and this one-dimensional situation was considered
in \ci{LS}. On the other hand, one may try to
identify KdV solitons with complex $k_i$ with a
special system of interacting electric and magnetic
charges \ci{Nien}.
We shall not consider the latter situation here,
but rather concentrate upon the purely electric systems on the
plane with real energy of interaction.

There are generalizations of the expressions
\re{iphase}-\re{iKDV_phase}
such that the corresponding $u(x, t, \dots)$ satisfy higher order
members of the KdV-hierarchy, $\sin$-Gordon,
Kadomtsev\,--\,Petviashvili (KP),
Toda, and some other integrable equations \cite{AS}. Note that
all these hierarchies can be derived within the free fermion
formalism which is outlined in the concluding section
of the present article.
In the next section we consider in detail an identification
of the KP hierarchy solitons and plasma particles.

\section{KP hierarchy}

The KP hierarchy which starts from the KP equation
$$
\frac{\partial^2 u}{\partial y^2}=\frac{\partial}{\partial x}
\left(\frac{\partial u}{\partial t}-
6u\frac{\partial u}{\partial x}+\frac{\partial^3 u}{\partial x^3}\right)
$$
is obtained through a generalization of the KdV formalism. The
KdV equation is an equation which describes the eigenvalue
preserving deformations of a second order differential operator,
while the KP equation describes that of a first order
pseudo-differential operator:
$$
L(t,\partial)\psi(t,k)=k\psi(t,k),
$$
$$
\frac{\partial}{\partial t_n}\psi(t,k)=B_n(t,\partial)\psi(t,k),
$$
where $t=(t_1=x,t_2=y,t_3=t,\dots), \quad\partial=\frac{\partial}{\partial t_1}$, and
$$
L(t,\partial)=\partial+u_1(t)\partial^{-1}+u_3(t)\partial^{-2}+\dots,
$$
\begin{equation}
B_n(t,\partial)=\partial^n+\sum_{j=0}^{n-1}b_{nj}(t)\partial^j.
\label{bnj}\end{equation}
The KP hierarchy is a system of nonlinear differential equations for $u_j(t)$ and
$b_{nj}(t)$, resulting from the Zakharov-Shabat type compatibility conditions
of linear equations given above
$$
\frac{\partial L}{\partial t_n}=\left[B_n,L\right] , \qquad
\frac{\partial B_m}{\partial t_n}-\frac{\partial B_n}{\partial
t_m}=\left[B_n,B_m\right].
$$
As in the KdV case, the potential $u=u_1$ is expressed in terms of the tau function
(\ref{pottau}) and $N$-soliton solution is presentable in the Hirota form (\ref{tau})
with the following parameterization
of the soliton phases and phase shifts \cite{DJKM}:
$$
A_{zz^\prime}=\ln\frac{(a_z-a_{z^\prime})(b_z-b_{z^\prime})}
{(a_z+b_{z^\prime})(b_z+a_{z^\prime})},
$$
\begin{equation}
\theta(z)=\theta^{(0)}(z)+\sum_{p=1}^\infty(a_z^p-(-b_z)^p)t_p,
\label{kp}
\end{equation}
where $t_p$ is the $p$-th KP ``time" and $a_z, b_z$ are some arbitrary
functions of $z$.

If we take
$$
a_z=z=x+\i y, \quad b_z=-z^*=-x + \i y, \quad y \ge 0
$$
then
\begin{equation}\lab{kpint}
A_{zz'}=W(z,z^\prime)=-2 V(z,z'),
\end{equation}
where
$$
V(z,z')=-\ln |z-z^\prime| + \ln |z^*-z^\prime|
$$
is the potential at the point $z$ created by a positive unit charge
particle placed at the point $z'$ over the conducting surface occupying
the $y\leq 0$ region (since $V(z,z')$ solves the Poisson equation with
the tangent boundary condition (\ref{tangent}) at $y=0$).
Using the method of images one may say that this is
an effective potential created by a positive charge at the point
$z^\prime, \; \Im z^\prime >0,$ and its image of opposite charge
located at $\left(z^\prime\right)^*$.
Comparing (\ref{kpint}) with the original definition (\ref{W})
one finds that the temperature (\ref{gamma}) is fixed and equals to
\begin{equation}
\Gamma=2.
\label{G1}
\end{equation}

Thus the reduction of KP $N$-soliton solution described above
corresponds to the Coulomb plasma in the upper half plane
$\Im z>0$ with metallic boundary along the $x$-axis.
This situation is depicted in the Fig. 1.
\begin{figure}
\begin{center}
\leavevmode
\epsfxsize = 111pt
\epsfysize = 158pt
\epsfbox{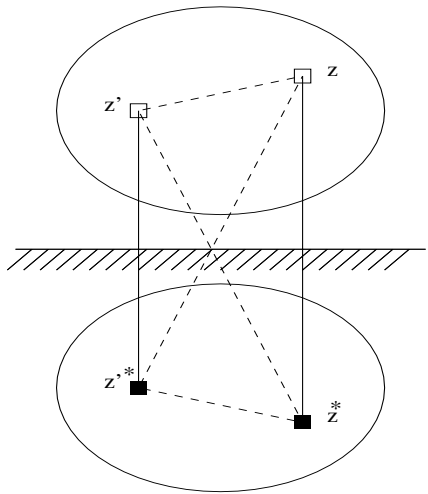}
\end{center}
\caption{KP equation: one component two-dimensional plasma above an
ideal conductor. Positive charges are shown as white squares while their negative
images are shown as black squares. Interactions between different
charges are shown by dashed lines, while the interactions between
charges and their own images are shown by the solid lines.}
\end{figure}

Note that the choice $a_z=z, b_z=z^*$ gives a similar situation ---
a plasma in the right half plane with the metallic boundary
along the $y$ axis (evidently this configuration is reached by
the rotation of the coordinates $z\to \i z$).

Let us shift $z\to z+\i a$, $a$ real and take the limit
$a\to\infty$, i.e. take the plasma far away from the boundary.
This leads to some divergences in the energy which can be removed by
addition of an appropriate diverging constant to the initial Hamiltonian.
As a result one gets the pure plasma system at the inverse temperature
$\Gamma=2$. This temperature corresponds to the standard normalization
in the random matrix theory \cite{G}. Note that the normalization
of the temperature in our previous papers \cite{LS} was chosen as
$\Gamma=1$ since there we were discussing Ising chains without
detailed comparison with the Coulomb systems which is a goal of
the present work.

The identification $w(z)=\theta(z)$ allows one to write the following
expression for the zero-time phases $\theta^{(0)}(z)$ in (\ref{kp})
\begin{equation}
\theta^{(0)}(z)=\mu-\Gamma(\ln |z^*-z|+ \phi(z)),\quad \Gamma=2,
\label{fkp}
\end{equation}
where the second term corresponds to the ``charge-image" self-interaction
energy, and the last term describes the potential of the
field created by the neutralizing background of the density $\rho(z)$:
\begin{equation}
\Delta\phi(z)=-2\pi\rho(z), \quad
\left.\frac{\partial \phi}{\partial x}\right|_{y=0}=0.
\label{rkp}
\end{equation}
Contributions from the KP ``times"
\begin{equation}
\sum_{p=0}^\infty (z^p-(z^*)^p)t_p=-\Gamma \phi_{ext}(z)
\label{z}
\end{equation}
correspond to an external electric field.
One can see that (\ref{z}) is a sum of the
harmonic polynomials, $p$-th polynomial being a fundamental
antisymmetric polynomial of
the group of reflections (the Coxeter group) of $2p$-gon.
Since the Laplacian of this part
is zero the corresponding density of charges is zero, i.e. this
field is generated by external distant charges.
For instance, the contribution of the first time $(z-z^*)t_1$
corresponds to the homogeneous electric field perpendicular to the
boundary. For reality of the potential the time $t_1$ has to be
purely imaginary.

The system of distant charges $g_i$ located at the points $w_i$
above the conductor create the following electrostatic potential
at the point $z$
$$
\phi_{ext}(z)=-\sum_i g_i\ln \frac{|z-w_i|}{|z-w^*_i|}.
$$
Since the external charges are far from the origin $| z | \ll | w_i |$,
we can expand the potential $\phi_{ext}(z)$ in the Taylor series and
get as KP times
$$
t_p\equiv \frac{\Gamma}{2p}\sum_i g_i\left(\frac{1}{(w_i^*)^p}
-\frac{1}{w_i^p}\right).
$$
In this picture the KP times take imaginary values automatically.

One may conclude that the general imaginary times evolution
of a special system of KP hierarchy solitons describes electrostatics
of a plasma in a varying external electric field.

It is worth of mentioning that sometimes the continuous limit is not
physical below some temperature.
Indeed, let us consider the partition function of $N$ particle
gas at the inverse temperature $\Gamma$ in a square domain
$ (0\le x=\Re z\le L) \times (0 \le y=\Im z \le L) $
$$
Z_N=\int_0^L{\rm d} x_1 \int_0^L {\rm d}y_1 \dots \int_0^L
{\rm d}x_N \int_0^L {\rm d} y_N
\exp\left(\Gamma\sum_{i<j}\ln\frac{|z_i-z_j|}{|z_i-z_j^*|}-\Gamma
\sum_i\ln|z_i-z^*_i|\right).
$$
Introducing new dimensionless variables $z_i/L$ one gets
the following expression for the thermodynamic potential or
the pressure $P$ in appropriate units
$$
P=\frac{\ln Z_N}{N\Gamma}=\frac{(2-\Gamma)\ln L}{\Gamma}
+ {\rm terms}\;{\rm independent}\; {\rm on}\; L.
$$
We see that at the inverse temperature $\Gamma=2$
the average pressure changes the sign.
The system is inhomogeneous and the local pressure is different
from the average one. It is clear that the local
pressure changes the sign at the boundary and at the inverse
temperatures $\Gamma\ge 2$ the particles stick to the
surface of the conductor.

As a result, the continuous space version of the
model in a domain with the conducting boundary
does not make physical sense if the temperature is
low enough.  However, the situation is different
if plasma is confined to a domain which does not touch
the boundary (as in the situation depicted in the
Fig. 1). E.g., one can avoid such an
unphysical behavior in the discrete (lattice) version of the model.
In the latter case the lattice
spacing plays the role of the radius of a hard core
repulsive interaction which prevents collapse of the
system. Note that the identification with the KP
soliton systems takes place exactly at the critical
temperature $\Gamma=2$.

Closing this section let us note that the
equation (\ref{laplace}) and the boundary conditions (\ref{normal}),
(\ref{tangent}) hold for an appropriate conformal change of the
variable $z\to f(z)$. For instance,
choosing soliton parameters in (\ref{kp}) as follows
$a_z=z^2, b_z=-(z^*)^2$, we obtain
$$
W(z, z')=2\ln\left|\frac{z^2-(z')^2}{(z^*)^2-(z')^2}\right|
$$
corresponding to the interaction of two charged particles in the rectangular
corner with metallic walls along the $x$ and $y$ axes. Higher
degree monomial maps $z\to z^n$ put the plasma into the
corner with the $\pi/n$ angle between the conducting walls.

The exponential map
\begin{equation}
a_z=\exp\left(\pi z/L\right), \quad b_z=-\exp\left(\pi z^*/L\right),
\label{strip}
\end{equation}
generates the $W$-potential
\begin{equation}
W(z,z^\prime)
=2\ln\left|\frac{\sinh\frac{\pi}{2L}(z-z^\prime)}
{\sinh\frac{\pi}{2L}(z^*-z^\prime)}\right|,
\label{potential}
\end{equation}
which describes the plasma in the strip $\Im z=(0,L)$
between two parallel conductors.

The choice
$$
a_z=\exp(-\pi x/L),\qquad b_z=-\exp(-\pi(x+\alpha)/L),
$$
results in
\begin{equation}
W(x,x')=\ln\frac{\sinh^2\frac{\pi(x-x')}{2L}}
{\sinh\frac{\pi}{2L}(x-x'-\alpha)\sinh\frac{\pi}{2L}(x-x'+\alpha)}.
\label{KPdip}\end{equation}
Solution of the Poisson equation with periodic boundary
condition along the $y$-axis with the period $2L$ is given by
the potential \cite{FJT}
$$
V(z,z')=-\ln \left|\sinh\frac{\pi}{2L}(z-z')\right|.
$$
Therefore one can interpret (\ref{KPdip}) as the interaction energy of two
neutral dipoles in the periodic background with the distance between
charges in the molecule equal to $\alpha$
(the internal energy of dipoles is neglected). These dipoles
all lie on the $x$-axis and have identical orientation.
Since $W=-\Gamma V$, we see that the effective inverse temperature is
twice smaller than in the previous cases, i.e. $\Gamma=1$.

Let us choose parameters in (\ref{kp}) as specific
double periodic functions
$$
a_z={\rm sn}^2\; z, \quad b_z=-{\rm sn}^2\; z^*,
$$
where ${\rm sn}\; z$ is the Jacobian elliptic function
with the periods $L_x, \i L_y$. Then one gets the $W$-potential
\begin{equation}\label{metbox}
W(z,z')=2\ln\left|\frac{ {\rm sn}^2\; z-{\rm sn}^2\; z'}
{{\rm sn}^2\; z^* - {\rm sn}^2\; z'}\right|
=2\ln\left|\frac{\theta_1(z-z')\theta_1(z+z')}{\theta_1(z^*-z')
\theta_1(z^*+z')}\right|,
\end{equation}
where $\theta_1(z)$ is the Jacobi theta-function
$$
\theta_1(z)= 2\sum_{n=0}^\infty (-1)^n \exp\left(-\pi
(n+1/2)^2\frac{L_y}{L_x}\right) \sin \frac{\pi(2n+1)z}{L_x}.
$$
It vanishes when $z$ lies on the boundary of a $L_x\times L_y$
rectangle, i.e. we have the plasma in a box
with the conducting walls ($\Gamma=2$).
For small $z,z'$ (or for the large size box $L_x, L_y\to \infty$)
one recovers plasma in the rectangular corner.

\section{KP hierarchy: the two component plasma case}

Let us consider the situation when lattice points $\{z\}$ consist of
two subsets $\{z_\pm\}$, $\{z\}=\{z_-\}\cup \{z_+\}$.
Choosing the following identification of parameters in (\ref{kp})
\begin{equation}
a_{z_-}=z_-, \quad b_{z_-}=-z_-^*, \quad a_{z_+}=z^*_+, \quad b_{z_+}=-z_+,
\label{two}
\end{equation}
we get a model of the two component plasma
(the plasma consisting of two species of opposite charges)
above an ideal conductor at the inverse temperature $\Gamma=2$:
$$
W(z_\pm,z^\prime_\pm)=2\ln|z_\pm-z^\prime_\pm|-2\ln |z^*_\pm-z^\prime_\pm|,
$$
$$
W(z_\pm,z^\prime_\mp)=2\ln|z^*_\pm-z^\prime_\mp|-2\ln |z_\pm-z^\prime_\mp|,
$$
$$
\theta(z_\pm)=\mu_\pm - 2\ln|z_\pm-z^*_\pm|\mp 2\phi(z_\pm).
$$
After the conformal transformations $z\to z^2, e^z,$ etc one gets
the two-component plasma in the metallic rectangular corner, a strip, etc.

Note that the bulk properties of the two component plasma are different
from the ones of the one component systems.
Actually, statistical mechanics of the one component plasma is not
stable without metallic boundary
or neutralizing background, since its particles tend toward
boundaries repelling each other with long range (logarithmic)
forces. In the two component case screening is possible if
the system is neutral and the temperature is
high enough. Simple scaling analysis shows that the pure two component
plasma undergoes a transition in the bulk at $\Gamma=2$ \cite{G}.
At this temperature an
association of opposite charges into molecules (neutralization of plasma)
takes place. Again, there is no such a transition in the
lattice case because of the hard-core repulsion at the lattice spacing
distance.

It is well known that the two component homogeneous plasma without
boundaries is a specific representation of the
quantum Sine-Gordon or Thirring model \cite{SS}.
The inverse temperature (\ref{G1}) corresponds to the free fermion
point, but the model is (in principle) integrable for any $\Gamma$.
The Sine-Gordon model requires
renormalization if the coupling constant exceeds a critical value.
The lattice spacing (in the case of the lattice plasma) is
equivalent to the introduction of a cutoff in the
corresponding field theory.
Thus it is natural to associate the case of two component lattice
plasma above the metallic boundary with some discretized boundary
Sine-Gordon or Thirring model.

\section{KP hierarchy: some reductions}

In this section we consider a number of one-dimensional reductions of the
KP hierarchy and some self-similar soliton solutions. We describe
here only the most popular integrable systems and do not cover
all possible cases and their Coulomb gas interpretations.

1) We begin with the reduction considered earlier in the literature \cite{G}.
In this case, plasma is restricted to the line $y=Y$. The $W$-potential
is translationally invariant and equals to
$$
W(x,x^\prime)=\ln\frac{(x-x')^2}{Y^2+(x-x^\prime)^2}.
$$
Self-interaction energies of charges with their own images
are constant and may be neglected.

2) Reduction to the KdV hierarchy. In this case particles move along
the vertical line $x=0$ and the $W$-potential is
\begin{equation}\lab{kdvphase}
W(y,y^\prime)=\ln \frac{(y-y^\prime)^2}{(y+y^\prime)^2}.
\end{equation}
Now a non-trivial self-interaction term $\propto \ln|2y|$ enters
the definition of soliton phases.

3) Discrete KdV hierarchy. 
The phase shifts have the form \cite{HDKDV}
$$
A_{zz^\prime}=\ln \left(\frac{\sinh\frac{1}{2}(z-z^\prime)}
{\sinh\frac{1}{2}(z+z^\prime)}\right)^2.
$$
Taking $z$ to be purely imaginary $ z=\i\alpha $ we get the following result
$$
W(\alpha,\alpha^\prime)=\ln\left(\frac{\sin\frac{1}{2}(\alpha-\alpha^\prime)}
{\sin\frac{1}{2}(\alpha+\alpha^\prime)}\right)^2,
$$
which corresponds to the plasma restricted to an arc with the center of
the corresponding circle at the conductor's surface.
The self-interaction energy is $-\ln|2\sin\alpha|$.
It is sufficient to put $z= e^{\i\alpha}$ in (\ref{kpint})
in order to get this system from the KP soliton solutions.

4) The reductions admitting both left and right moving
solitons correspond to the plasma restricted to domains
of disjoint parts. E.g., the Toda lattice case \cite{AS}
$$
A_{zz'}=\ln\left(\frac{\epsilon(z) z - \epsilon(z^\prime)
z^\prime}{1-\epsilon(z)\epsilon(z^\prime) z
z^\prime}\right)^2,\quad \epsilon(z) =\pm 1
$$
corresponds for $z=e^{\i\alpha}$ to the two component
plasma on the half circle with the center lying
upon the conductor surface. The positive charge particles
occupy the subsector $\alpha\in [0,\pi/2]$ and the
negative charge particles are located in the subsector
$\alpha\in[\pi/2,\pi]$.

5) The Boussinesq  equation \cite{HB} corresponds to a one component plasma
on disjoint halves of the hyperbola $3x^2=y^2+1$ situated above
the conductor occupying the $y\leq 0$ half plane:
$$
A_{yy'}=\ln\frac{(\epsilon(y)x(y)-\epsilon(y^\prime)x(y^\prime))^2+(
y-y^\prime)^2}{(\epsilon(y)x(y)-\epsilon(y^\prime)x(y^\prime
))^2+(y+y^\prime)^2}, \quad x(y)=\sqrt{\frac{y^2+1}{3}}, \quad
\epsilon(y)=\pm 1.
$$

It would be interesting
to find the equation whose soliton phase shifts are given
by the elliptic functions \re{metbox}.

Consider some specific lattice plasma configurations
associated with self-similar soliton solutions of integrable equations.

Let us take the KP
hierarchy with parameters (\ref{strip}) and restrict the corresponding
plasma to the line parallel to the conductor surfaces
$$
\Im z=Y.
$$
Applying this constraint, we get from (\ref{potential})
$$
W(x-x^\prime)=-\ln\left(\sin^2\frac{\pi Y}{L}\coth^2\frac{\pi(x-x^\prime)}
{2L} + \cos^2\frac{\pi Y}{L}\right).
$$
The simplest self-similar reduction corresponds to the homogeneous
one-dimensional lattice parallel to the $x$-axis.
Then the $n$-th charge has the coordinates $z_n=X+hn+\i Y$, where
$X$ is some fixed constant and $h$ is the lattice spacing.
In this case soliton momenta $a_z$ and $b_z$ form one geometric
progression (\ref{strip}). The simplest KdV self-similar reduction
corresponds to the case when the line is located at equal distances
between parallel conductors $Y=L/2$:
\begin{equation}
W(x-x^\prime)=\ln\tanh^2\frac{\pi(x-x^\prime)}{2L}.
\label{line}
\end{equation}

$M$-periodic reduction corresponds to the situation
when soliton momenta are composed from $M$ distinct
geometric progressions.
In this case the lattice $\{ z\}$ consists
of $M$ homogeneous sublattices.  In the plasma language it
corresponds to the model where plasma moves on $M$ distinct
parallel lines between two conductors with the coordinates
$\{z\}=\{\i Y_p+ X_p+nh, p=1, \dots, M,\; n=0, 1, \dots\}$.
If one sets $Y_p=L/2$ then all these sublattices are situated
upon the middle line and this case corresponds to the general
self-similar KdV soliton potentials of \cite{S1}.

In a similar way one can place plasma upon $M$ parallel
lines with fixed $x$-coordinates, $x=X_p$, which are
situated above the $y\leq 0$ conductor. In this case self-similar
lattices are described by restriction of $y$ to the union of $M$
geometric progressions $\{z\}=\{X_p+\i Y_pq^n, q<1\}$.
However, such configurations are not safe from the collapse
of particles on the walls. The KP self-similar
systems are richer than the ones of the KdV or BKP (to be considered below)
cases due to the presence of non-trivial translational parts
$X_p$ (or $Y_p$) in the parameterization of the corresponding soliton
spectral data $a_z, b_z$.

\section{Correspondence with Ising lattices}

It is well known that lattice gas models are related to the Ising
models \cite{B}. Indeed, substituting
$$
\sigma(z)=\left(s(z)+1\right)/2, \quad s(z)=\pm 1,
$$
into (\ref{G}), we get a formula for the partition function of an
Ising model up to some constant multiplicative factor
\begin{equation}
G=\sum_{\{s\}}\exp\left(\frac{1}{2} \sum_{z\neq z^\prime}
J\left(z,z^\prime\right)s(z)s(z^\prime)+\sum_{\{z\}}H(z)s(z)\right),
\label{IG}
\end{equation}
where
$$
J(z,z^\prime)=\frac{1}{4}W(z,z^\prime)
$$
are the exchange constants and
\begin{equation}
H(z)=\frac{1}{2}w(z)+\frac{1}{4}\sum_{z^\prime, z^\prime \neq z}W(z,z^\prime)
\label{H}
\end{equation}
denotes the external magnetic field. Here we have absorbed the
temperature variable $\Gamma$ into the definition of $J(z,z')$
and $H(z)$.

Note that the Ising models emerging in this way look
natural only in the reduced one-dimensional case, when there are no
constraints upon the values of spins at different points. In the
two-dimensional picture with boundaries one has to assume that
the configuration of spins satisfies some geometric constraints
which do not have natural meaning similar to the one existing in
the electrostatics. This leads to the exchange which depends not
only on the distance between the spins but on the distance
to the boundaries as well.

Technically, it appears that the Ising representation of the lattice plasma
grand partition
function is more convenient for writing it in the determinant form.

\section{BKP hierarchy}

Consider the BKP equation:
$$
{\partial \over \partial t_1} \left( 9{\partial u \over \partial t_5}
-5 {\partial^3 u \over \partial t_3 \partial t_1^2}+{\partial^5 u \over
\partial t_1^5} - 30{\partial u \over \partial t_3}{\partial u \over
\partial t_1}+
30{\partial u \over \partial t_1}{\partial^3 u \over \partial t_1^3}+
60\left({\partial u \over \partial t_1}\right)^3 \right)
 - 5{\partial^2 u \over \partial t_3^2}=0.
$$
This equation involves three independent variables $t_1, t_3, t_5$.
It is the first equation of the B-type KP (BKP) hierarchy which is a
reduction of the general KP hierarchy \cite{DJKM,JM}.
This reduction is achieved by setting $b_{n0}=0$ in (\ref{bnj}),
which assumes, in turn, that even ``times" evolution is absent, $t_{2p}=0$.
The BKP tau-function is defined via the relation $u(t_1, t_3, \dots)=
\partial \ln\tau_{BKP}/\partial t_1$.

$N$-soliton solution of the BKP hierarchy has the same form (\ref{tau})
with the following phase shifts and soliton phases \cite{DJKM}:
$$
A_{zz^\prime}=\ln\frac{(a_z-a_{z^\prime})(b_z-b_{z^\prime})
(a_z-b_{z^\prime})(b_z-a_{z^\prime})}{(a_z+a_{z^\prime})(b_z+b_{z^\prime})
(a_z+b_{z^\prime})(b_z+a_{z^\prime})}, \quad
$$
\begin{equation}
\theta(z)=\theta^{(0)}(z)+\sum_{p=1}^\infty(a_z^{2p-1}+b_z^{2p-1})t_{2p-1},
\label{bkp}
\end{equation}
where $t_{2p-1}$ are the BKP ``times".

\begin{figure}
\begin{center}
\leavevmode
\epsfxsize = 235pt
\epsfysize = 220pt
\epsfbox{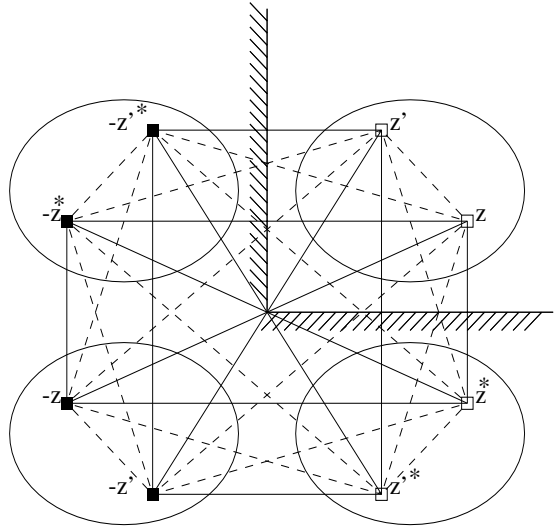}
\end{center}
\caption{BKP equation: a one component two-dimensional plasma
in the corner between an ideal
dielectric (the horizontal axis) and an ideal conductor (the vertical axis).
Positive charges are shown as white squares while their negative charge
images are shown as the black squares. Interactions between different
charges are shown
by dashed lines, while the self-interactions between charges and their
own images are shown by the solid lines.}
\end{figure}

If we take one component plasma  and fix
$$
a_z=z=x+\i y, \quad b_z=z^*=x-\i y, \quad x>0,\quad y \ge 0,
$$
then $A_{zz'}=W(z,z')= -2V(z,z'),$ where
$$
V(z,z^\prime)=-\ln|z-z^\prime|-
\ln|z^*-z^\prime|+\ln|z+z^\prime|+\ln|z^*+z^\prime|
$$
is the total interaction energy between unit charge particles at the points
$z, \; \Im z >0, \Re z > 0,$ and $z^\prime, \; \Im z^\prime >0,
 \Re z^\prime >0,$
in a domain of the upper right quarter of the
plane with an ideal dielectric boundary (cf. (\ref{normal}))
along the $x$-axis and an ideal
conductor wall along the $y$-axis (cf. (\ref{tangent})).
This situation is depicted in the Fig. 2.
According to (\ref{gamma}) the inverse temperature is fixed and equals to
$ \Gamma=2.$

If the plasma is taken far away from the corner by
appropriate translations, then one
gets the pure plasma system at the effective temperature $\Gamma=2$.
Sliding along the $y$-axis to infinity one comes to the previously
considered plasma associated with the KP equation.
Sliding along the $x$-axis requires a renormalization of the
zero energy level after which one gets a plasma above the surface of a
dielectric.

If we place charges upon the $y=0$ axis, then the dielectric
boundary condition disappears and we have
$$
W(x,x')=\ln\left(\frac{x-x'}{x+x'}\right)^4,
$$
which corresponds to the plasma model induced by the
KdV equation at the inverse
temperature $\Gamma=4$, which is twice higher than in
the appropriate KP reduction case.

Identification of the initial phase $\theta^{(0)}(z)$ in (\ref{kp}) is
as follows
$$
\theta^{(0)}(z)=2\ln|z^*-z|-2\ln|z^*+z|-2\ln|2z|+\mu - 2\phi(z),
$$
where the meaning of all terms is similar to that of
(\ref{fkp}), (\ref{rkp}). Contribution of the BKP ``times"
\begin{equation}
\sum_{p=1}^\infty (z^{2p-1}+(z^*)^{2p-1})t_{2p-1}
\label{Z}
\end{equation}
corresponds to electric field created by external charges.
The potential (\ref{Z}) satisfies all
the boundary conditions (\ref{normal}), (\ref{tangent}).
The $p$-th polynomial in the sum
(\ref{Z}) is the fundamental antisymmetric polynomial of the
reflection group of the $2(2p-1)$-gon.
Again, the evolution of solitons under
the BKP flow corresponds to the evolution of plasma under a
motion of distant external charges.

In complete parallel with the KP case one can consider
conformal transformations $z\to z^n, e^z,$ etc and map
plasma to various geometric configurations.
Considering systems with two sublattices like (\ref{two})
one arrives at the model of two component plasma in the
metal-dielectric corner or other bounded regions.
This leads again to a boundary Sine-Gordon model at the free
fermion point corresponding to the temperature $\Gamma=2$.

\section{BKP hierarchy: a dipole gas on a line between two ideal conductors}

In this section we consider real self-similar reductions of the
BKP hierarchy corresponding to dipole gases on a line between two conductors.

Let us write out once more the $N$-soliton $\tau$-function of the BKP
hierarchy in the slightly different notations
$$
\tau_N=\sum_{\sigma=0,1}\exp\left(\sum_{1\le i< j\le N}A_{ij}\sigma_i\sigma_j
+\sum_{i=1}^N\theta_i\sigma_i\right),
$$
$$
e^{A_{ij}}=\frac{(a_i-a_j)(b_i-b_j)(a_i-b_j)(b_i-a_j)}{(a_i+a_j)
(b_i+b_j)(a_i+b_j)(b_i+a_j)},
$$
where $i,j$ are the numbers of solitons (they are equivalent in meaning
to the coordinate $z$ of plasma particles).
Choosing
\begin{equation}
a_i=\exp(-\pi h i/L),\qquad b_i=-\exp(-\pi(h i+\alpha)/L), \quad i=1,\dots, N,
\label{ab}
\end{equation}
we get an expression for the grand partition function of the form
(\ref{G}) for a homogeneous lattice gas.
The particles of the gas interact via the following $W$-potential
\begin{equation}
W_d(i-j)=
W(h(i-j))-\frac{1}{2}W\left(h(i-j)-\alpha\right)-\frac{1}{2}W\left(h(i-j)+\alpha\right)
\label{www}
\end{equation}
with $W$ given by (\ref{line}). It is seen immediately that this is
the potential of two dipole molecules
consisting of two opposite charges situated at the distance $\alpha$
from each other without the part describing interaction of
charges inside the dipoles (which is constant). The distance between
molecules at $i$-th and $j$-th site is $h(i-j)$, where $h$ denotes
the lattice spacing. The dipoles move along
the line situated at equal distances $1/2L$ from
two parallel conductors (see the Fig. 3). All dipoles are
oriented in one direction. Due to the dipole gas interpretation
one has the effective temperature $\Gamma=1$ --- this is
a demonstration of some discrete temperature renormalization effect.

If one takes $a_i$ as in (\ref{ab}) but changes the sign of
$b_i$ (i.e. shifts $\alpha\to\alpha+\i L$), then the signs
of the second and third terms in (\ref{www}) are changed too.
This situation corresponds to a dipole gas on the middle line,
the dipoles being charged molecules of total charge +2 with the same
distance between charges $\alpha$. These molecules are positioned
similar to the previous case.

Another lattice gas model of charged
dipoles appears if one replaces $\alpha$ in (\ref{www})
by $\i(\alpha+L), \; 0<\alpha< L.$ In this case charged
dipoles are positioned vertically and symmetrically with respect
to the middle line $y=L/2$.  The case when $\alpha$ in (\ref{www})
is replaced by $\i\alpha, 0<\alpha< L$, corresponds to the
neutral dipoles gas in the strip between {\it dielectric} walls.
The dipoles are perpendicular to the middle line $y=L/2$, similar
to the previous case, and have an identical orientation.

The $M$-periodic self-similar reductions \cite{LS,S1},
when $a_{i+M}=qa_i, b_{i+M}=qb_i$, describe the gas 
consisting of $M$ different types of dipoles.
In some particular cases this leads to dipole gases with
different orientations of neutral or charged molecules in a strip
between the conducting walls.

For instance, for the choice
$$
a_{2i}=e^{-\frac{\pi}{L}(2ih-\alpha/2)},\qquad b_{2i}=-e^{-\frac{\pi}{L}(2ih+\alpha/2)},\quad
$$
$$
a_{2i+1}=e^{-\frac{\pi}{L}((2i+1)h+\alpha/2)},\qquad
b_{2i+1}=-e^{-\frac{\pi}{L}((2i+1)h-\alpha/2)}
$$
neutral dipoles situated on the even and odd sites
have opposite directions. (Here one may note that the formal substitution
$\sigma_i \to (-1)^i\sigma_i +1-(-1)^i$ in the grand partition function
converts the interaction energy between molecules to the previous
form (\ref{www}). However, this transformation changes the form of
interaction with external fields.)

The $2M$-periodic reduction
$$
\begin{array}{lll}
a_{2jM-M}=e^{-\frac{\pi}{L}\left(2jh-\i\alpha_M\right)}
&\quad&
b_{2jM-M}=-e^{-\frac{\pi}{L}\left(2jh+\i\alpha_M\right)} \\
a_{2jM-M+1}=e^{-\frac{\pi}{L}\left(2jh-\i\alpha_{M-1}\right)}
&\quad&
b_{2jM-M+1}=-e^{-\frac{\pi}{L}\left(2jh+\i\alpha_{M-1}\right)} \\
\dots &\quad& \dots \\
a_{2jM-1}=e^{-\frac{\pi}{L}\left(2jh-\i\alpha_1\right)}&&
b_{2jM-1}=-e^{-\frac{\pi}{L}\left(2jh+\i\alpha_1\right)} \\
a_{2jM}=e^{-\frac{\pi}{L}\left((2j+1)h+\i\alpha_1\right)}  & \quad &
b_{2jM}=-e^{-\frac{\pi}{L}\left((2j+1)h-\i\alpha_1\right)} \\
\dots &\quad& \dots \\
a_{2jM+M-2}=e^{-\frac{\pi}{L}\left((2j+1)h+\i\alpha_{M-1}\right)}
&\quad&b_{ 2jM + M -2 } = -e^{-\frac{\pi}{L}\left((2j+1)h -\i\alpha_{M-1}\right)}\\
a_{2jM+M-1}=e^{-\frac{\pi}{L}\left((2j+1)h+\i\alpha_M\right)}&
\quad &
b_{2jM+M-1}=-e^{-\frac{\pi}{L}\left((2j+1)h-\i\alpha_M\right)}
\end{array} $$
describes a gas of neutral dipoles lying on the middle line between two ideal dielectrics.
Dipoles are pointed normally to the boundaries. They can switch
their orientations (``up" and ``down")
and internal degrees of freedom characterized by $M$ different
dipole moments $\alpha_j.$ In general, we have to introduce $M$ different chemical
potentials describing ``internal energy" of the dipole molecule.

Another possible generalization describes mixtures of the $+2$ charge molecules
with both parallel and perpendicular orientations of dipoles.
Such models can describe polar plasmas where molecules can perform
discrete rotations by $\pi/2$. More complicated types of mixtures
of plasma particles are possible as well.
Several physical variables can be calculated here. These are the number
density, polarization and the pressure. One particular model is solved
in the next section.

\begin{figure}
\begin{center}
\leavevmode
\epsfxsize = 222pt
\epsfysize = 200pt
\epsfbox{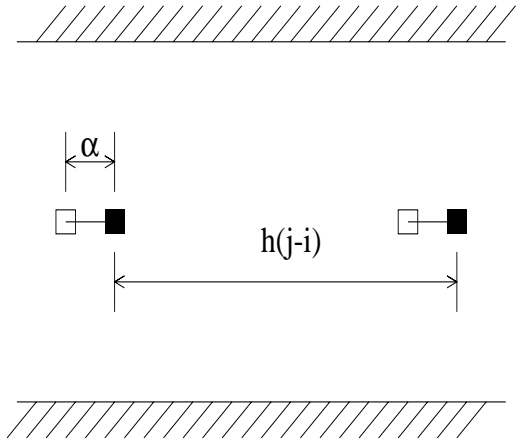}
\end{center}
\caption{BKP: dipole gas on the line between two ideal conductors.
The dipoles are neutral, oriented identically and lie upon
the homogeneous lattice.}
\end{figure}

\section{Solution of a dipole gas model}

It is known from the theory of solitons that the $\tau$-function (\ref{tau})
can be represented as a determinant of some matrix. In general it is
necessary to apply the inverse scattering
method for an auxiliary linear problem to write solutions in such a form.

Determinant representations are of great help for evaluations of
the partition functions,
since the corresponding matrices appear to have the Toeplitz form in some physically
interesting cases and, as a result, they can be diagonalized by the discrete
Fourier transformation. Here we consider only the BKP case, since
some models related to the KP hierarchy happen to be
considered already in the literature, see, e.g. \cite{G}.

As shown in \cite{LS}, the $N=2p$ soliton BKP $\tau$-function,
or the corresponding Ising model partition function (\ref{IG})
admits the following determinant form depending on the
soliton parameters $a(z), b(z)$ and the magnetic field (\ref{H}):
\begin{equation}
G_{2p}=
\left(\prod_{z \neq z^\prime}
\frac
{(a(z)+a(z^\prime))(b(z)+b(z^\prime))(a(z)-b(z^\prime))(b(z)-a(z^\prime))}
{(a(z)-a(z^\prime))(b(z)-b(z^\prime))(a(z)+b(z^\prime))(b(z)+a(z^\prime))}
\right)^{1/8}\sqrt{\det {\rm\bf Gr}},
\label{BKP_pfaffian}
\end{equation}
where matrix elements of the matrix {\bf Gr} are
$$
{\rm\bf Gr}_{z,z^\prime} =
g(z)g(z^\prime)\frac{b(z)-b(z^\prime)}{b(z)+b(z^\prime)}e^{H(z)+H(z^\prime)}
+g(z)g(z^\prime)^{-1}\frac{b(z)+a(z^\prime)}{b(z)-a(z^\prime)}
e^{H(z)-H(z^\prime)}
$$
$$
+g(z)^{-1}g(z^\prime)\frac{a(z)+b(z^\prime)}{a(z)-b(z^\prime)}
e^{-H(z)+H(z^\prime)}
+g(z)^{-1}g(z^\prime)^{-1}\frac{a(z)-a(z^\prime)}{a(z)+a(z^\prime)}
e^{-H(z)-H(z^\prime)},
$$
and
\begin{equation}
g(z)=
\left(\prod_{z^\prime \atop z^\prime \ne z}\frac{(a(z)-a(z^\prime))(b(z)+b(z^\prime))
(b(z)-a(z^\prime))(a(z)+b(z^\prime))}
{(a(z)+a(z^\prime))(b(z)-b(z^\prime))(b(z)+a(z^\prime))(a(z)-b(z^\prime))}\right)
^{1/4}.
\label{BKP_coefficients}
\end{equation}

We consider the situation when the external potential is constant,
$w=\mu$, i.e. the dipoles interact only between themselves.

Since the potential (\ref{www}) is of the short range type,
the magnetic field in the
corresponding Ising model (\ref{H}) is homogeneous in the bulk.
We can neglect the field
inhomogeneities at the edges in the thermodynamic limit $N\to\infty$.
>From (\ref{H}) it follows that
$$
H=\frac{1}{2}(\mu+C), \quad C=\frac{1}{2}\sum_{i=1}^\infty W_d(i).
$$

Consider the dipole gas model described in the Fig. 3.
Substituting (\ref{ab}) into (\ref{BKP_coefficients}) we see that
the matrix ${\rm\bf Gr}_{ij}$
has the Toeplitz form, ${\rm \bf Gr}_{ij}={\rm\bf Gr}_{i-j}$,
in the thermodynamic limit $p\to\infty$. Therefore it is diagonalized
by the discrete Fourier transformation. The final answer for the
thermodynamic potential per site or the pressure in appropriate
units $P(\mu)=\lim_{p\to\infty}(\ln G)/2p$ can be found from the
results of calculations of the free energy per site of
the corresponding Ising chain \cite{LS}:
\begin{equation}
P(\mu)=\frac{1}{4}\ln\frac{(q, q, bq, q/b; q)_\infty} {(-q,
-q, -bq, -q/b; q)_\infty} +\frac{1}{4\pi}\int_0^{2\pi}d\nu \ln|2\rho(\nu)|,
\label{BKPpress}\end{equation}
where $q=e^{-\pi h/L}$, $b=-e^{-\pi\alpha/L},$ and
$$ \rho(\nu)=\cosh (C+\mu)
+\frac{(-q;q)_\infty^2}{(-e^{\i\nu}, -qe^{-\i\nu}; q)_\infty}
\left(
\frac{(b^{-1}e^{\i\nu}, qbe^{-\i\nu}; q)_\infty}{(b^{-1}, qb; q)_\infty}
+\frac{(be^{\i\nu}, qb^{-1}e^{-\i\nu}; q)_\infty}{(b, qb^{-1}; q)_\infty}
\right).
$$
The standard notations for the $q$-shifted factorials
$$
(a; q)_0=1,\quad (a; q)_n=(1-a)(1-aq)\cdots (1-aq^{n-1}),
$$
$$
(a_1, a_2, \dots, a_n; q)_n=(a_1; q)_n(a_2; q)_n\dots (a_n;q)_n
$$
are used in these formulae.

Taking the derivative with respect to $\mu$
we find the number density of molecules $n(\mu)$:
\begin{equation}
n(\mu)=\frac{1}{2}+\frac{1}{2}
\left(1-\frac{1}{\pi}\int_0^\pi \frac{d\nu}
{1+d(\nu)\cosh (\mu+C)}\right) \tanh (C+\mu),
\label{magbkpgen}
\end{equation}
where
$$
d(\nu)=\frac{(qb,q/b;q)_\infty (|b|^{-1/2}+|b|^{1/2})
\theta_2(\nu, q^{1/2})}
{(-q;q)_\infty^2 2\mbox{Re }\theta_2(\nu-(\i/2)\ln|b|, q^{1/2})}.
$$
Here $\theta_2(\nu, q^{1/2})$ is the Jacobi theta-function
$$
\theta_2(\nu, q^{1/2})=2\sum_{n=0}^\infty q^{(n+1/2)^2/2}\cos{(2n+1)\nu}
$$
$$
=2q^{1/8}\cos\nu \; (q;q)_\infty
(-qe^{2i\nu};q)_\infty(-qe^{-2i\nu};q)_\infty.
$$
In a similar way one can analyze other dipole gas models, in
particular, the general $M$-periodic reduction cases, mentioned
in the previous section.

In conclusion of the discussion of the BKP hierarchy it should be
mentioned that statistical mechanics interpretation of the KP reductions
of the C and D types (CKP and DKP) \cite{JM} is not known to the
authors. One may try to find a Coulomb gas realization of soliton
solutions of these and other known integrable equations which were
not considered in this paper.

\section{Conclusions}

Constructions considered so far have one essential drawback from the point
of view of statistical physics: the partition functions derived from the
tau functions of classical integrable hierarchies correspond only to
some fixed temperatures. A possible way of overcoming this obstacle is
to look for appropriate quantum generalizations of classical hierarchies.

It is well known (e.g., see \cite{SS}) that the
grand partition function of two component plasma can be
expressed in terms of the following equivalent field theories:
Sine-Gordon, Thirring or sigma-model.

Lattice versions of the neutral Coulomb plasma correspond to some
discretizations of the above models: lattice Sine-Gordon, scalar Hubbard
or XXZ model. Since these three models are equivalent, we take the
scalar Hubbard model as their representative.

The two component plasma at the temperature (\ref{G1}) is mapped onto
the free fermion point of the one-dimensional scalar Hubbard model.
It can also be mapped to free spin 1/2
fermions on the lattice \cite{CJ}. The
Coulomb plasma at the temperatures different from (\ref{G1}) corresponds
to interacting fermions.

>From our point of view, the relation of Coulomb plasmas to the
theory of integrable hierarchies described here sheds some new light onto the
relation between fermion models and different kinds of plasmas.
Indeed, it is known \cite{JM} that the soliton  equations can be derived in
the framework of the free fermion formalism and they are equivalent to
Plucker relations on the infinite dimensional Grassmanian manifold.
We remind briefly basic points of this formalism.

Let us take fermions on the one-dimensional lattice. Their creation and annihilation
operators $\psi^*_i, \psi_i$ satisfy the relations:
$$
\{\psi^*_i, \psi_j\} = \delta_{ij},
\qquad \{\psi_i, \psi_j\} = 0, \quad i,j \in {\bf Z}.
$$
The neutral vacuum is defined as the state where the sites with $i<0$ are
empty and other sites are filled by fermions (actually, it is a kink
state of the XX model or the free fermion point
of the scalar Hubbard model).

The action of the fermion operators on the vacuum is
$$
\psi_n\vert {\rm vac} \rangle =0,\; n<0,
\qquad \psi_n^*\vert {\rm vac} \rangle =0,\; n\ge 0,
$$
$$
\langle {\rm vac} \vert \psi_n=0, \; n \ge 0,
\qquad \langle {\rm vac} \vert \psi_n^*=0, \; n < 0.
$$
The fermion field operators
$$
\psi(\lambda)=\sum_{k=-\infty}^\infty\psi_k\lambda^k,\quad
\psi^*(\lambda)=\sum_{k=-\infty}^\infty\psi_k^*\lambda^{-k}
$$
evolve under the KP flow as follows
$$
e^{H(t)}\psi(\lambda)e^{-H(t)}=e^{\theta(t,\lambda)}\psi(\lambda), \quad
e^{H(t)}\psi^*(\lambda)e^{-H(t)}
=e^{-\theta(t,\lambda)}\psi^*(\lambda),
$$
$$
\theta(t,\lambda)=\sum_{n=1}^\infty t_n\lambda^n,
$$
where the generating function of KP Hamiltonians $H(t)$ has the form
\begin{equation}
H(t)=\sum^\infty_{n=1} H_n t_n, \quad
H_n=\sum_{k=-\infty}^\infty:\psi_k\psi_{k+n}^*:,
\label{hhhh}\end{equation}
where the colons mean the normal ordering with respect to
the vacuum $|vac\rangle$.
The $\tau$-function is defined as follows
$$
\tau(t,g)=\langle{\rm vac} \vert e^{H(t)} g \vert{\rm vac} \rangle,
\quad g \in G,
$$
where $G$ is a subgroup of the fermion algebra preserving one-fermion states
$$
G=\left\{g\vert \; g {\cal V} g^{-1} = {\cal V},\; g {\cal V}^* g^{-1} = {\cal V}^*\right\}
$$
and ${\cal V} = \oplus_{i \in {\rm Z}} C \psi_i,\; {\cal V}^* =
\oplus_{i \in {\rm Z}} C \psi_i^*.$
Choosing the element $g$ of $G$ in the definition of $\tau$-function as
$$
g=\exp \sum_{i=1}^N \gamma_i \psi(a_i)\psi^*(-b_i),
$$
where $\gamma_i$ are some arbitrary constants related to
the zero time phases of solitons, one gets the KP
$N$-soliton $\tau$-function (\ref{tau}), (\ref{kp}).

Thus, the two component plasma at the temperature (\ref{G1}) is described
simultaneously by the KP hierarchy and the free fermions model. Variation of
the temperature from this value, which in some models (e.g., in the
one-dimensional Ising chains picture) is not qualitatively distinguished from
the other ones, would correspond to a generalization of the free fermion
formalism for integrable hierarchies to the case of interacting fermions.
The KP hamiltonians which are of the form of
hopping terms of the Hubbard model, suggest that such a generalization is
possible, provided KP hamiltonians are replaced by conserved quantities
of a generalization of the spinless Hubbard or XX model.
For instance, in the XX limit of the XXZ model
derivatives of the transfer matrix are in the algebra (\ref{hhhh})
$$
(\ln T(u))^{(n)} \propto H_{-n}+H_n.
$$
One may conjecture that the derivatives of general XXZ transfer
matrix belong to a generalization of the algebra of free fermion
Hamiltonians of the KP hierarchy.

Here we encounter one difficulty. Roughly speaking, the algebra
of the KP Hamiltonians (\ref{hhhh}) has two times as many
Hamiltonians as derivatives of the XXZ transfer matrix.
Probably there are some extra integrals of motion
which are reduced to $H_n-H_{-n}$ in the free fermion
limit. As an example, we can mention that some extra nonlocal constants
of motion have been derived for the Hubbard model in \cite{BSS}.

Another possible way of generalization is due to the approach described,
e.g., in \cite{IIKS},
where some generalization of the nonlinear Schr\"odinger equation
is presented. Unfortunately the
notion of $\tau$-function is clearly defined only in the limiting cases of
the free fermion points (e.g., for the impenetrable Bose gas).
It is not obvious that solutions of the corresponding integro-differential
equations can be expressed in terms of a $\tau$-function at general
couplings and that such a function would make sense in the plasma picture.

Concluding this article we discuss possible physical significance of the
models considered in it going beyond the Coulomb gas picture.
As was shown in \cite{LS} there are nice interpretations of the
one-dimensional cases from the point of view of Ising magnets. However,
various boundary conditions arising within the intrinsically two-dimensional
Coulomb interaction systems look somewhat artificial in the Ising picture.
Still, a number of 2D Ising models with such non-local
exchange can be formulated which are exactly solvable by the
techniques due to Gaudin (see the third paper in \cite{G}).

Another possible application concerns the
fractional quantum Hall effect (FQHE).
Remind that the two-dimensional one component Coulomb plasma in the uniform
neutralizing background of the density $1/2m$
has the following partition function
$$
Z=\int{\rm d} z_1\dots {\rm d} z_N
\exp\left(-2mE\right), \quad
E=-\sum_{i<j}\log|z_i-z_j| + \frac{1}{4m}\sum_i|z_i|^2.
$$
It was shown by Laughlin that the $n$-point correlation functions
in the appropriate state of the FQHE at the filling factor $1/m$ coincide
with those of the one component Coulomb plasma. In particular, the
normalization factor of the corresponding wave function is given by $Z$.

Our model is a bit different from the pure one component plasma,
since we have non-trivial boundary conditions. However, by placing
the domain of concentration of the charged particles far from the
boundary we get the Laughlin plasma. It is also possible that
the plasma with boundaries has some meaning in the fractional quantum
Hall picture as well describing there some boundary effects.
$\tau$-functions of the classical hierarchies
correspond to boundary Laughlin states at the full filling $1/m=1$.
Nontrivial examples $m>1$ could be described by $\tau$-functions
of the generalized hierarchies discussed above.

{\it Acknowledgments.} The authors are indebted to F.D.M.
Haldane, V.I. Inozemtsev, V.E. Korepin, S.P. Novikov and
V.B. Priezzhev for some remarks and stimulating discussions.
The work of I.L. is supported by a fellowship from NSERC (Canada),
V.S. is supported in part by the RFBR (Russia) grant 97-01-01041 and
the INTAS grant 96-0700.

\end{document}